\documentclass[11pt,twoside]{article}

\usepackage{asp2006_hotcool}
\usepackage{graphicx}
\usepackage{amssymb}

\begin{document}
\title{Tepid Supergiants: Chemical Signatures of Stellar Evolution \& The
Extent of Blue Loops}   
\author{Norbert Przybilla$^1$, Markus Firnstein$^1$, and Mar\'ia-Fernanda Nieva$^2$}
\affil{$^1$\,Dr. Remeis Observatory, Sternwartstr. 7, D-96049 Bamberg, Germany\\
$^2$\,MPI for Astrophysics, Postfach 1317, D-85741 Garching, Germany}

\begin{abstract}
Massive stars can develop into tepid supergiants at several stages of their
post main-sequence evolution, prior to core He-burning, on a blue loop, or  
close to the final supernova explosion. 
We discuss observational constraints on models of massive star evolution
obtained from the analysis of a sample of Galactic supergiants and put them
in the context of the cosmic abundance standard as recently proposed from
the study of their OB-type progenitors ($Z$\,$=$\,0.014 for stars in
the solar neighbourhood). High-precision abundance analyses for 
He and CNO, with uncertainties as low as $\sim$10-20\%, trace the transport efficiency of
nuclear-processed material to the stellar surface, either by rotational
mixing or during the first dredge-up. A mixing efficiency higher by 
a factor $\sim$2 than predicted by current evolution models for rotating stars
is indicated, implying that additional effects need to be considered in
evolutionary models like e.g. the interplay of circulation and magnetic
fields. Blue loops are suggested to extend to higher masses and
to higher $T_{\rm eff}$ than predicted by the current generation of
stellar evolution models.
\end{abstract}

\section{Introduction}
Massive stars of $\sim$8-40\,$M_{\odot}$ spend a short time in between the 
hot and cool extremes of the Hertzsprung-Russell diagram (HRD) as tepid supergiants 
(BA-type SGs, $T_{\rm eff}$\,$\sim$\,8000-15\,000\,K). Possible scenarios are the
first crossing from blue to red after the main sequence phase, blue loop
episodes or a final crossing from red to blue in a late evolution stage
close to the final supernova explosion. 

Important observational 
constraints on the evolution of massive stars can be derived from studies of 
tepid SGs via the determination of stellar parameters and abundance patterns
for light elements.
In particular, enriched helium and nitrogen, and depleted
carbon are indicative for the mixing of nuclear-processed matter
from the stellar core to the atmosphere. Recent models of massive star
evolution accounting for mass loss and rotation
\citep{hela00,mame00} and the interplay of rotation and magnetic fields 
\citep{heger05,mame05} provide detailed predictions on the expected effects
of chemical mixing on surface abundances.
Systematic investigations of massive stars of different mass and
metallicity can provide the necessary observational constraints to
distinguish between competing treatments of the complex
(magneto)hydrodynamic processes, which are not fully understood from first
principles. These processes redistribute angular momentum, transport 
nuclear-processed products from the core to the surface and replenish
hydrogen, thus extending lifetimes and increasing the stellar luminosity.

Here we report results from {\em precision analyses} of Galactic tepid
supergiants and a control sample of unevolved B-type stars in the solar
neighbourhood, i.e. the progenitors of BA-type SGs. The derived observational
constraints are confronted with present-day model predictions and the conclusions from
this may guide future refinements to stellar evolution models.

\section{Observations \& Quantitative Spectral Analysis}
High-S/N and high-resolution Echelle spectra
($S/N$$>$200,\,$R$\,=\,$\lambda/\Delta\lambda$\,$\approx$\,40-48\,000)
of Galactic BA-SGs and unevolved B-type stars were obtained using FEROS on the
ESO La Silla 1.5m/2.2m and FOCES on the Calar Alto 2.2m telescopes.
The wide wavelength coverage ($\sim$3900-9200\,{\AA}) provided access to 
all spectroscopic indicators required for the quantitative analysis.

The model calculations were carried out in a hybrid non-LTE approach as
discussed in detail by \citet{prz06} and \citet{nipr07}. In brief, 
hydrostatic, plane-parallel and
line-blanketed LTE model atmospheres were computed with {\sc Atlas9}
(Kurucz~1993; with further modifications: Przybilla, Butler \& Kudritzki~2001) and non-LTE 
line formation was performed on the resulting model stratifications with 
{\sc Detail} and {\sc Surface} (Giddings~1981; Butler \& Giddings~1985;
both updated by K. Butler). The former solves the coupled 
radiative transfer and statistical equilibrium equations while
the latter performs the formal solution with refined line-broadening theories. 
State-of-the-art non-LTE model atoms relying on data from {\em ab-initio} 
computations -- avoiding rough approximations wherever possible -- were utilised.

The stellar parameter and abundance determination for BA-SGs and unevolved B
stars relied on the iterative analysis methodology described by \citet{prz06} and
\citet{nipr07,nipr08}. Numerous spectroscopic indicators like multiple
ionization equilibria and Stark-broadened profiles of the higher Balmer and Paschen 
lines were simultaneously used to derive effective temperatures $T_{\rm eff}$
and surface gravities $\log g$. The accuracy of the method allows the 
1$\sigma$-uncertainties to be reduced to $\sim$1-2\% in $T_{\rm eff}$  and to 
0.05-0.10\,dex in $\log g$, when the atmospheric helium content, metallicity
and microturbulence are correctly accounted for in a self-consistent way. 
Absolute elemental abundances can then be constrained with unprecedented
accuracy ($\sim$10--20\%, random, and $\sim$25\%, systematic
1$\sigma$-errors; see also Nieva \& Przybilla, these proceedings).

\section{Initial Conditions: A Cosmic Abundance Standard}
It is well established that metallicity plays an important r\^ole in the evolution
of massive stars \citep[e.g.][]{mame00}. Most evolution models
representative for the massive star population in the solar neighbourhood
assume a value of (solar) metal mass fraction of $Z$\,$=$\,0.02.

Slowly-rotating unevolved early B-type stars were long since recognised as best
targets for elemental abundance determinations at present day, as
they show the simplest photospheres among the massive stars. They even allow
pristine abundances for helium, carbon and nitrogen to be derived. All previous
studies of early B-type stars in the solar neighbourhood found a wide
scatter in abundances -- and therefore metallicity --, favouring overall
sub-solar values, see e.g. \citet{przybilla08}.  

\begin{table}[!t]
\caption{\hspace{-4.2mm}Cosmic abundance standard: light elements \& global mass fractions}
\label{cas}
\smallskip
\begin{center}
{\small
\begin{tabular}{llll}
\tableline
He: 10.98\,$\pm$\,0.02 & C: 8.35\,$\pm$\,0.05 & N: 7.76\,$\pm$\,0.05 & O: 8.76\,$\pm$\,0.03\\
$X$\,$=$\,0.715    & $Y$\,$=$\,0.271  & $Z$\,$=$\,0.014\\
\tableline
\end{tabular}
}
\end{center}
\end{table}

In the meantime, improvements in model atmosphere analyses of cool stars
lead to a revision of the solar abundances \citep*{asplund05}. A wide
range of systematic effects on quantitative analyses of early B stars have
also been identified (Nieva \& Przybilla, these proceedings), providing a
sound basis for a re-investigation of elemental abundances in these stars. 

This re-investigation indicates that elemental abundances of early B stars
in the solar neighbourhood are in fact highly homogeneous, showing a rms
scatter as low as $\sim$10\,\% \citep*{prz08}, the same as reported for ISM
gas-phase abundances. A {\em cosmic abundance standard} for the present-day solar
neighbourhood can thus be proposed, with light element abundances (given as
log\,(El/H)\,$+$\,12) and mass
fractions for hydrogen ($X$), helium ($Y$) and metals ($Z$) as summarised in Table~\ref{cas}.
Here, pristine carbon abundances are adopted from \citet{nipr08}. In
general, good agreement with the revised solar abundances \citep{asplund05}
is found, with some small differences in oxygen (and neon).
These cosmic abundances are recommended as initial values for future stellar
evolution calculations of massive stars. Note in particular the reduction of $Z$ to 0.014
from the so far ca\-nonical value of 0.02.

\section{Observational Constraints on Massive Star Evolution}
Accurate atmospheric parameters and elemental abundances have been
determined for 15 Galactic tepid supergiants so far
\citep{prz06,firnstein06,schipr08}. Six BA-type supergiants are located 
within Gould's Belt (like all B star targets), while the remaining ones are
found in the field and in associations out to distances of $\sim$3\,kpc 
from the Sun. 

All sample stars within Gould's Belt show similar abundances of the heavier
elements \citep{prz08}, corroborating the concept of
chemical homogeneity of the ISM in the solar neighbourhood at present day. A
slightly larger spread in abundances is found for the more distant stars,
implying a total range in abundance of less than a factor of 2
\citep[and flat Galactic abundance gradients, Figs.~18 \& 19 in][]{przybilla08}, which is 
consistent with predictions from simple models of turbulent mixing within the
ISM \citep[e.g.][]{roy95}. This is much less than the factor of $\sim$10 in
abundance range found in previous studies.  

\begin{figure}[!t]
\begin{center}
\includegraphics[width=0.64\linewidth]{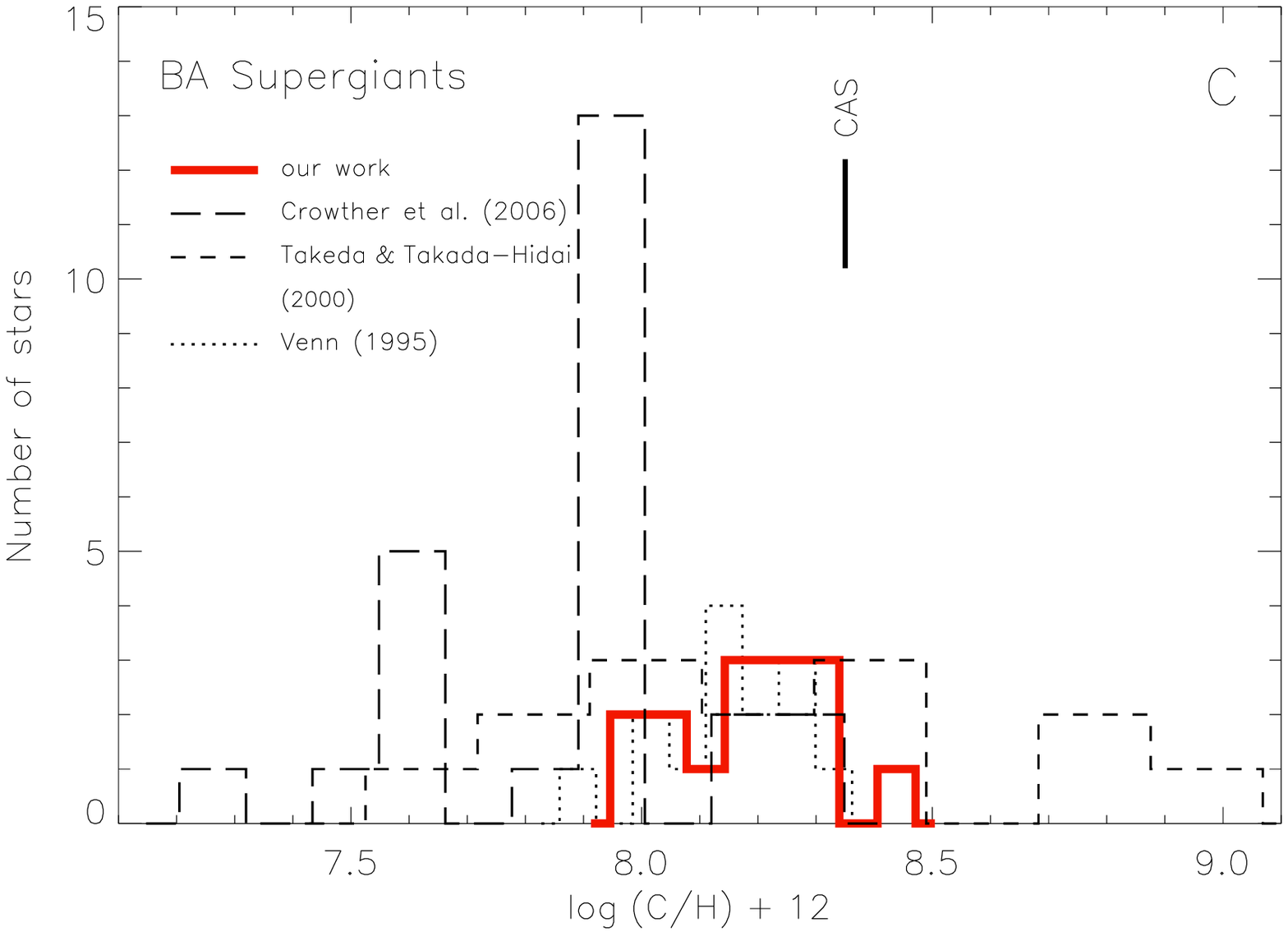}\\
\includegraphics[width=0.64\linewidth]{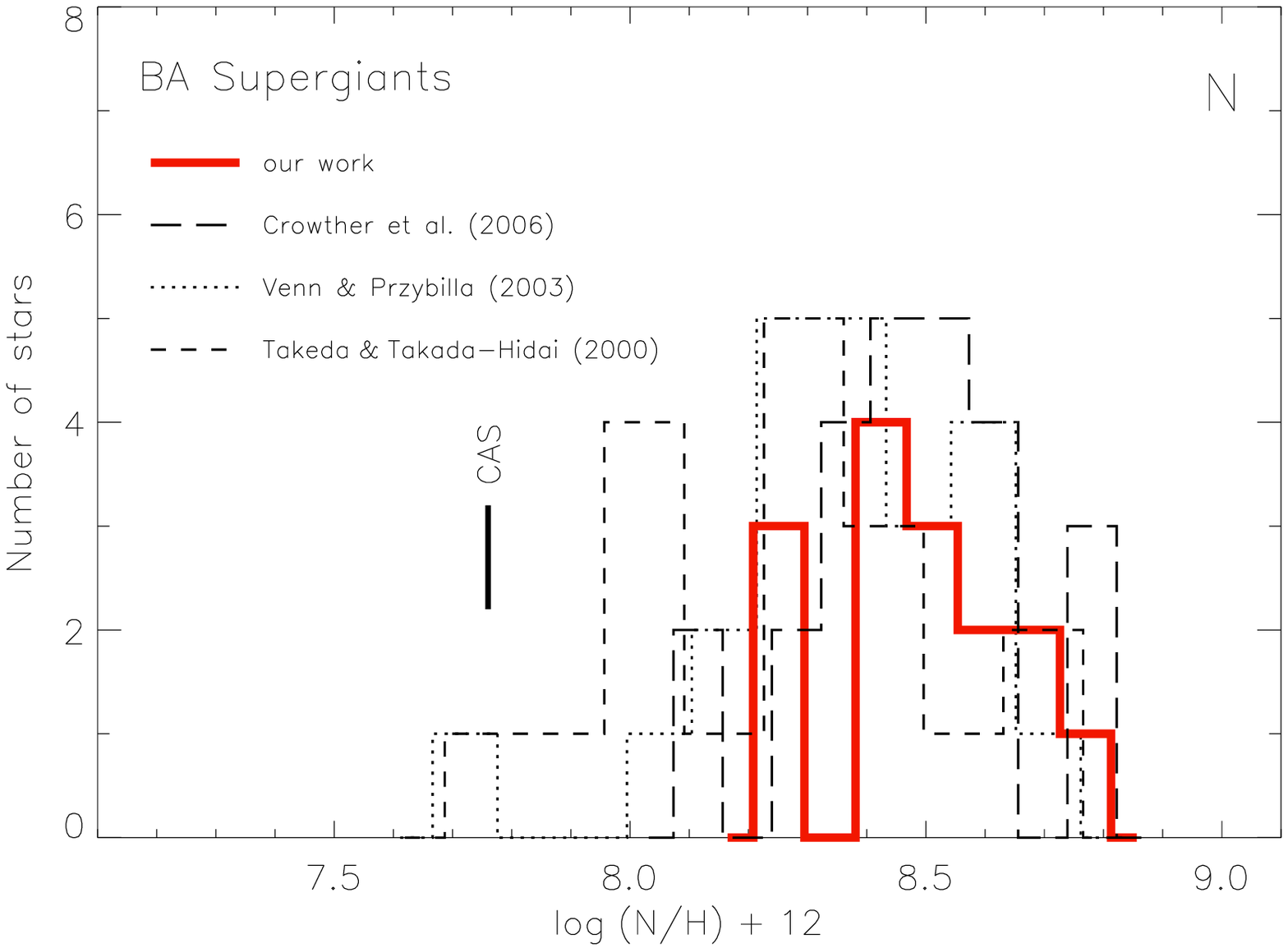}\vspace{-7mm}
\end{center}
\caption{Comparison of carbon and nitrogen abundances in BA-type supergiants 
as obtained in our work and from the literature. Cosmic standard abundances
(CAS) from Table~\ref{cas} are indicated by vertical lines.
Bin width is $\sigma$/2 of the individual studies, with $\sigma$ denoting the
rms scatter.}
\label{histograms}
\end{figure}

Data relevant for our comparison with stellar evolution models are
displayed in Fig.~\ref{histograms}. The abundance distributions show
depleted carbon and enriched nitrogen relative to the cosmic abundance
standard values, as expected for the evolved state of the
supergiants. The upper limit of carbon abundances in the BA-type supergiants is consistent
with the pristine value derived from the B stars, meeting a boundary
condition imposed by mixing. Note that the abundance ranges covered are much smaller 
than in published work on similar stars (as indicated in Fig.~\ref{histograms}) -- a
consequence of the largely reduced systematic uncertainties in our analysis,
which applies also to the findings for other elements.

\begin{figure}[!t]
\includegraphics[width=0.85\linewidth]{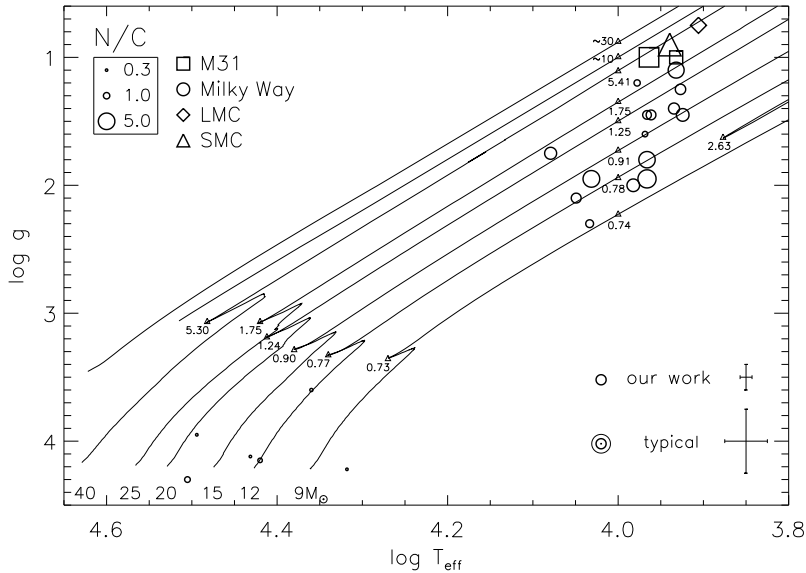}\vspace{-5mm}
\caption{
Observational constraints on massive star evolution.
Displayed are results for the most sensitive indicator for mixing with
nuclear-processed matter, N/C, in a homogeneously analysed sample of
Galactic BA-type supergiants (circles) and their progenitors on the main
sequence, B stars. The N/C ratios are
encoded on a logarithmic scale, with some examples indicated.
Error bars (1$\sigma$-statistical \& systematic) from
our work and those typical for previous work are also indicated.
Stellar evolution tracks \citep{mema03} are
displayed for rotating stars at $Z_\odot$\,$=$\,0.02 (full lines). 
Starting with an initial N/C$\sim$0.3,
theory predicts N/C$\sim$1 for BA-type supergiants evolving to the red, and
N/C$\sim$2--3 after the first dredge-up (markers along the tracks). Since we 
find N/C values as high 
as $\gtrsim$6 the observed mixing efficiency is higher than
predicted. Also, blue loops extend to hotter temperatures than predicted. 
Results for a few objects in nearby galaxies are also shown.
}
\label{evol}
\end{figure}

The comparison of observed N/C abundance ratios with predictions from stellar
evolution models is made in Fig.~\ref{evol} (preliminary results for a few
highly massive supergiants in nearby galaxies are also shown). We recall that 
this comparison is not 
ideal, as the evolution calculations \citep{mema03} were performed for $Z$\,$=$\,0.02, 
while the observations are on average consistent with a value of $Z$\,$=$\,0.014. 
However, the expected effects for this
metallicity difference will not change the general conclusions drawn in the
following. 

Most of the apparently slow-rotating 
B stars show N/C ratios close to the pristine value of $\sim$0.3. A notable exception is
$\tau$\,Sco, which was shown recently to be a truly slow-rotating magnetic star.
The situation is more complex with the BA-type supergiants. They all
exhibit slow rotation because of their expanded envelopes, irrespective of
the initial
rotational velocity of their progenitors on the main sequence. It appears
that the objects at masses
below $\sim$15\,$M_\odot$ show larger amounts of nuclear-processed material
than those around $\sim$20\,$M_\odot$. Larger N/C ratios are found again
for the most massive objects of the sample. Note that objects of
$M$\,$\gtrsim$\,30\,$M_\odot$ are located either in the Magellanic Clouds or
in M31, i.e. they have a different metallicity than the Galactic stars.
Note further that in view of the results discussed by Levesque et al. (these
proceedings) stars of such high masses apparently avoid the red
supergiant phase.  

Two important conclusions can already be drawn from this small sample.
Let us concentrate first on the objects more massive than $\sim$15\,$M_\odot$.
A general trend of increased mixing of nuclear-processed material with
increasing stellar mass is found, in
accordance with the predictions of evolution models. Moreover, the strongest mixing
signature is found for the most metal-poor object, AzV\,475 in the Small
Magellanic Cloud, also in agreement with theory \citep[e.g.][]{mame01}.
However, the mixing efficiency appears to be higher
(by a factor of $\sim$2) than predicted by current state-of-the-art evolution
computations for rotating stars with mass-loss. Stellar evolution models
accounting for the interplay of rotation and magnetic fields may resolve
this discrepancy since they find a higher efficiency for chemical
mixing \citep{mame05}.

Larger N/C ratios at $M$\,$\lesssim$\,15\,$M_\odot$ can be explained
if the objects are on a blue loop, i.e. if they have already undergone the
first dredge-up during a previous phase as a red supergiant, in addition
to rotational mixing. This interpretation is in contrast to earlier findings
of \citet{venn95}.
Further support for the blue-loop scenario comes from lifetime considerations.
Stellar evolution calculations indicate that supergiants spend a much longer
time on a blue-loop (with core He-burning) than required for the
crossing of the HRD from the blue to the red (the short
phase of core contraction after core H-burning has ceased).
E.g., in the case of a rotating 9\,$M_\odot$ model of \citet{mema03}
the difference is about a factor of 15. It is well-established that blue loops
are required to explain the Cepheid variables, but their extent in the HRD
-- in particular the upper limits in temperature and stellar mass -- are
essentially unknown. The blue-loop phase is highly sensitive to the details of
the stellar evolution calculations \citep[`{\ldots} is a sort of
magnifying glass, revealing relentlessly the faults of calculations of
earlier phases.',][]{kiwei90}.

Consequently, a systematic study of a larger sample of massive stars
could provide the tight observational constraints required for a thorough
verification and refinement of the stellar evolution models. Precision analyses of stars
covering the relevant part of the HRD are under way.

\acknowledgements 
M.\,F. and N.\,P. gratefully acknowledge financial support of the project by the Deutsche 
Forschungsgemeinschaft (grant PR\,685/3-1).

\end{document}